\documentclass[11pt]{article}

\begin{document}
\def\doublespaced{\baselineskip=\normalbaselineskip\multiply\baselineskip
 by 150\divide\baselineskip by 100}
\doublespaced
\def\lsim{~{\rlap{\lower 3.5pt\hbox{$\mathchar\sim$}}\raise 1pt\hbox{$<$}}\,}
\def\gsim{~{\rlap{\lower 3.5pt\hbox{$\mathchar\sim$}}\raise 1pt\hbox{$>$}}\,}
\def\thisday{~September~10, 1998 ~and~ hep-th/yymmnnn~~}
\def\thisday{~\today ~and~ hep-pth/yymmnnn~~}


\vbox to 3.5cm {
\vfill
}
\begin{center}{\Large{Transformation Property of the Caputo Fractional Differential Operator in Two Dimensional Space}}\\
Ehab Malkawi  \\
{{Department of Physics,\\
United Arab Emirates University\\
 Al Ain, UAE}}\\
\end{center}
\vskip -10pt

\begin{center}
{Abstract}
\end{center}
The transformation property of the Caputo fractional derivative operator of a scalar function under  rotation in two dimensional space is derived. The study of the transformation property is essential for the formulation of fractional calculus in multi-dimensional space. The inclusion of fractional calculus in the Lagrangian and Hamiltonian dynamics relies on such transformation. An illustrative example is given.

\vskip 6pt



\vskip 3pt

\newpage

\section{Introduction}

Fractional calculus deals with differentiation and integration to arbitrary noninteger orders,
which can be real or complex. The subject is by no means new. The idea appeared in a letter by
Leibniz to L'Hospital in 1695.  Full mathematical aspects and applications of fractional calculus can be found in Refs.~[1]-[4].

In recent years, fractional calculus has played a very important role in various fields such as mechanics, electricity, chemistry, biology, economics, control theory, signal image processing, and groundwater problems. However, most of the physics applications have been restricted to one-dimensional space problems. The inclusion of fractional
multi-dimensional space operators has not been studied thoroughly. A touch on this subject is discussed in Ref.~[5]. The study of the transformation properties of the fractional derivatives under space rotation can strongly shed light on the interpretation and manipulation of fractional derivatives in multi-dimensional space. Obviously, it is expected that fractional derivatives operators to behave differently from scalar, vector, and tensor objects because of their nonlocality. It is our aim in this paper to investigate such behavior in a simple form.

To employ fractional calculus into Lagrangian dynamics, it is essential to study the
behavior of partial derivatives under coordinate transformations.
For example, given a scalar function $\Phi(x,y)$ describing some physical
potential, the field $\nabla \Phi(x,y)$, determined by differentiation of
the scalar field $\Phi(x,y)$, is a vector field. To include the fractional derivatives
in the Lagrangian formulation, it is essential to investigate the transformation properties
of the fractional derivative operator $\nabla^\nu \Phi(x,y)$.

In this paper and using the Caputo definition we investigate the transformation properties of the fractional derivatives of a scalar function under
space rotation in two dimensions. An earlier work was done using the Riemann-Liouville definition in Ref.[6], however, we believe that the Riemann-Liouville definition is not suitable since the derivative of constant is not Zero. Thus the invariance of a constant scalar field is broken.  In section 2, we give a brief introduction to fractional calculus.
In section 3 we investigate the transformation properties of the fractional derivatives under two-dimensional coordinates rotation. Finally, in section 4 we give a brief discussion and conclusions.

\section{Fractional Calculus}

 Several definitions of the fractional differentiation and integration exist in literature. The most common used are the Riemann-Liouville and the Caputo derivatives.
The Riemann-Liouville derivative of a constant is not zero while Caputo’s derivative of a constant is zero. This property makes the Caputo definition more suitable in our work.  If we consider a constant scalar function $\phi(x,y)=C$ throughout a region in the $x,y$ plane, then the Caputo definition guarantees that the scalar function remains invariant in the rotated $x^\prime, y^\prime$ plane.
 The Caputo differential operator of fractional calculus is defined as
\begin{eqnarray}
  _aD_x^\alpha f(x)   \equiv   & & \frac{1}{\Gamma(n-\alpha)} \int_a^x {(x-u)}^{n-\alpha-1} \frac{d^n f(u)}{du^n}du\, , \,\,
   n-1 < \alpha < n  \nonumber \\
        & & \frac{d^n}{dx^n} f(x) \,  ,\,\,\,\,\,\,\,\,\,\,\,\,\,\,\ \,\,\,\,\,\,\,\,\,\,\,\,\,\,\,\,\,\,\,\,\,\,\,\,\,\,\,\,\,\ \,\,\,\,\,\,\,\,\,\,\,\,\,\,\,\,\,\,\ \,\,\,\,\ \alpha=n
\end{eqnarray}
where $\Gamma(.)$ is the Gamma function and $x>a$.
In this work we consider the case $a=0$, $0< \alpha<1$, $n=1$.
For the power function $x^p$, the Caputo fractional derivative satisfies
\begin{eqnarray}
D_x^\alpha x^p = & & \frac{\Gamma(p+1)}{\Gamma(p-\alpha+1)} x^{p-\alpha}\,\,\,\,  ,  0<\alpha<1, p>0, p\in R\, \nonumber \\
                           & &  0\,\,\,\,\, , \,\,\,\,\,\,\,\,\,\,\,\,\,\,\,\,\,\,\,\,\,\,\,\,\,\,\,\,\,\,\,\,\,\,\,\,\,\,\,\,\,\,\,  0<\alpha <1, p=0
\label{eq2}
\end{eqnarray}
For mathematical properties of fractional derivatives and integrals one can consult Ref.[1]-[4].

\section{Transformation of the Caputo Fractional Derivative of a Scalar Field Under Space Rotation}

In this work we only consider infinitesimal space rotation in two dimension $(x,y)$. The general case of three dimensions
will be left for future work.
Consider a scalar field $\Phi(x,y)$ which is required to be analytic at the point $(0,0)$.
The scalar function $\Phi(x,y)$ can be expanded in a Taylor series written as
\begin{eqnarray}
\Phi(x,y)=\sum_{n,m} a_{n,m}
\frac{x^n y^m}{n! m!}\, ,
\label{eq4}
\end{eqnarray}
where
\begin{eqnarray}
a_{n,m}\equiv \frac{\partial^{n}\partial^{m}\Phi(0,0)}{\partial x^n \partial y^n}
\end{eqnarray}

Under an infinitesimal rotation $\delta \theta$ in the $x-y$ plane, the coordinates $x$ and $y$
transform as
\begin{eqnarray}
x^\prime & = & x +\delta\theta y \nonumber \, , \\
y^\prime & = & y - \delta\theta x\, .
\label{xy}
\end{eqnarray}
The scalar function $\Phi(x,y)$
is invariant under space rotation, i.e., $\Phi^\prime(x^\prime,y^\prime)=\Phi(x,y)$.
To guarantee that $\phi^\prime(x^\prime,y^\prime)=\phi(x,y)$ one can show that, to first order in $\delta\theta$,
\begin{equation}
a^\prime_{n,m}= a_{n,m}+\delta\theta \left[ n a_{n-1,m+1} -ma_{n+1,m-1}\right]\, , \forall n,m .
\label{tr2}
\end{equation}
The partial derivatives of the scalar function transform as
\begin{eqnarray}
\frac{\partial}{\partial x^\prime}\,\phi^\prime(x^\prime,y^\prime) & = &
\frac{\partial}{\partial x} \phi(x,y) +\delta\theta \frac{\partial}{\partial y}\phi(x,y)\, ,\nonumber \\
\frac{\partial}{\partial y^\prime}\Phi^\prime(x^\prime,y^\prime)   & = &
\frac{\partial}{\partial y}\Phi(x,y) -\delta\theta \frac{\partial}{\partial x}\Phi(x,y)\, ,
\label{grad}
\end{eqnarray}
The transformations of the partial derivatives imply that the gradient
 $\nabla\Phi(x,y)$ is a vector field.

In this work we investigate the transformation properties of the Caputo fractional derivatives
$ D_x^\nu \Phi(x,y)$ and $D_y^\nu \Phi(x,y)$, where
$0\leq \nu\leq 1$. The end-point case $\nu=0$ refers to the identity transformation
$\partial^0/\partial x^0 \Phi(x,y)\equiv \Phi(x,y)$. The case $\nu=1$ refers to the usual first-order partial derivatives. The reason to choose the Caputo definition is to guarantee that a constant scalar function remains invariant under space rotation, since $D^\nu c=0$, where $c$ is constant.

We start by investigating the transformation property of the fractional partial derivative
$D_x^\nu\Phi(x,y)$. Starting with Eq.~(\ref{eq4}) and using the result of
Eq.~(\ref{eq2}) we write the fractional derivative as
\begin{eqnarray}
D_x^\nu\Phi(x,y) & = &
\sum_{n\geq 1 ,m}  a_{n,m} \frac{x^{n-\nu} y^m}{n! m!}
\frac{\Gamma(n+1)}{\Gamma(n-\nu+1)} \nonumber \\
 & = &
\sum_{n\geq 1,m} a_{n,m} \frac{x^{n-\nu} y^m}
{\Gamma(n-\nu+1) m!}\, .
\end{eqnarray}
Similarly, the transformed partial derivative can be written as
\begin{equation}
D^{\nu}_{x^\prime} \Phi^\prime(x^\prime,y^\prime) =
\sum_{n\geq 1,m} a^\prime_{n,m}
\frac{x^{\prime(n-\nu)} y^{\prime m}}{\Gamma(n-\nu+1)! m!}\,\, .
\label{trans}
\end{equation}
Using the transformations in Eq.~(\ref{xy})
we can write, to first order in $\delta\theta$,
\begin{eqnarray}
x^{\prime{n-\nu}} & = & {\left(x+\delta\theta \, y\right)}^{n-\nu}=
x^{n-\nu} +(n-\nu)\, \delta\theta\, x^{n-\nu-1}\, y \nonumber \,, \\
y^{\prime{m}} & = & {\left(y-\delta\theta \, x\right)}^{m}=
y^{m} - m\, \delta\theta\, x\, y^{m-1} \, .
\label{tr1}
\end{eqnarray}

Substituting the results of Eqs.~(\ref{tr1}) and (\ref{tr2}) into Eq.~(\ref{trans}) one finds
\begin{eqnarray}
  D^\nu_{x^\prime} \Phi^\prime(x^\prime,y^\prime)
  &=& \sum_{n\geq 1,m} \frac{1}{m!\Gamma(n-\nu+1)}
  [a_{n,m} +\delta\theta\, n\, a_{n-1,m+1} -\delta\theta \,m\, a_{n+1,m-1}]  \nonumber \\
   & &  [x^{n-\nu} y^m +(n-\nu) \delta\theta x^{n-\nu-1}y^{m+1} - m\delta\theta x^{n-\nu+1}y^{m-1}]\, .
\end{eqnarray}

Keeping terms to first order in $\delta\theta$ we find
\begin{eqnarray}
  D^\nu_{x^\prime} \Phi(x^\prime,y^\prime
    &=&  \sum_{n\geq 1,m} \frac{1}{m!\Gamma(n-\nu+1)}[a_{n,m} x^{n-\nu} y^m  \nonumber \\
    & + & \delta\theta\, \{ (n-\nu)\,a_{n,m} x^{n-\nu-1}y^{m+1}  \nonumber \\
   & -&  m \, a_{n,m} x^{n-\nu+1} y^{m-1} \nonumber \\
    &+ & n\, a_{n-1,m+1} x^{n-\nu} y^m  \nonumber \\
    & - & m\, a_{n+1,m-1} x^{n-\nu} y^m\} ]\, .
\end{eqnarray}
By shifting powers of $x$ and $y$ cancellation occurs and the above equation reduces to
\begin{equation}
D^\nu_{x^\prime} \Phi^\prime(x^\prime,y^\prime)=
D_x^\nu \Phi(x,y)+
\nu \delta\theta
\sum_{n,m} \frac{1}{\Gamma(n-\nu+1) m!} a_{n-1,m+1} x^{n-\nu}y^m\, .
\end{equation}
The last term can be rewritten as
\begin{equation}
\nu \delta\theta D_x^\nu I_x^1 D_y^1 \phi(x,y)\, ,
\end{equation}
where $D_y^1\phi(x,y)=\partial \phi(x,y)/{\partial y}$ and $I_x^1\phi(x,y)=\int_0^x \phi(x,y)dx$.
Thus we write the final result as
\begin{equation}
D^\nu_{x^\prime} \Phi^\prime(x^\prime,y^\prime)=
D_x^\nu \Phi(x,y)+\nu \delta\theta D_x^\nu I_x^1 D_y^1 \phi(x,y)\, .
\end{equation}
Similarly we can show that
\begin{equation}
D^\nu_{y^\prime} \Phi^\prime(x^\prime,y^\prime)=
D_y^\nu \Phi(x,y)-\nu \delta\theta D_y^\nu I_y^1 D_x^1 \phi(x,y)\, .
\end{equation}
The above transformation property of the fractional derivatives reflects the inherited non-locality of the fractional calculus. A small variation in the fractional derivatives of a scalar function at the point $(x,y)$ in a coordinate system is connected to the evolution (integral) of the function in another rotated coordinate system.

\section{Discussion and Conclusions}
For the special case $\nu=0$, it is straightforward to recover the invariance
\begin{equation}
\Phi^\prime(x^\prime,y^\prime)=\Phi(x,y)\, .
\end{equation}
For the case $\nu=1$, we recover the vector
transformations in Eq.~(\ref{grad}) since $D^1_xI_x^1=1$.

As an example, consider the scalar field $\Phi(x,y)=x^2+y^2$.
It is straightforward to derive the transformation of the partial derivatives
\begin{equation}
D^\nu_{x^\prime}\Phi^\prime(x^\prime,y^\prime)=
D_x^\nu\Phi(x,y)+\frac{2\delta\theta}{\Gamma(2-\nu)}   x^{1-\nu}y\, ,
\end{equation}
and
\begin{equation}
D^\nu_{y^\prime} \Phi^\prime(x^\prime,y^\prime)=
D_y^\nu\Phi(x,y) -\frac{2\delta\theta}{\Gamma(2-\nu)} x\,y^{1-\nu}\, .
\end{equation}
It becomes a simple exercise to show that the quantity
\begin{equation}
x^\nu \frac{\partial^\nu\Phi(x,y)}{\partial x^{\nu}} +
y^\nu \frac{\partial^\nu\Phi(x,y)}{\partial y^{\nu}}\,.
\end{equation}
is a scalar, i.e., invariant under space rotation.
Also one can show that for the above function the fractional partial operator
$(D^\nu_x D^\nu_x + D^\nu_y D^\nu_y )\Phi(x,y)$ is not invariant.
However, the quantity $( x^{2\nu}\,D^\nu_x D^\nu_x + y^{2\nu}\,D^\nu_y D^\nu_y )\Phi(x,y)$ is invariant in two dimensional space.  Therefore, there is a need to fully study the behavioral transformations of the fractional partial derivatives before a serious attempt to generalize fractional calculus into the three-dimensional space. Further future studies are needed to understand this behavior.

\section*{Acknowledments}
The author would like to thank Prof. Yuri Luchko for useful discussion.

\section*{References}

\begin{itemize}
  \item[[1]]
   Fractional Integrals and Derivatives: Theory and Applications, by Samko, S.; Kilbas, A.A.; and Marichev, O. Hardcover: 1006 pages. Publisher: Taylor \& Francis Books. ISBN 2-88124-864-0

\item[[2]]
An Introduction to the Fractional Calculus and Fractional Differential Equations, by Kenneth S. Miller, Bertram Ross (Editor). Hardcover: 384 pages. Publisher: John Wiley \& Sons; 1 edition (May 19, 1993). ISBN 0-471-58884-9

\item[[3]]
The Fractional Calculus; Theory and Applications of Differentiation and Integration to Arbitrary Order (Mathematics in Science and Engineering, V), by Keith B. Oldham, Jerome Spanier. Hardcover. Publisher: Academic Press; (November 1974). ISBN 0-12-525550-0

\item[[4]]
Fractional Differential Equations. An Introduction to Fractional Derivatives, Fractional Differential Equations, Some Methods of Their Solution and Some of Their Applications., (Mathematics in Science and Engineering, vol. 198), by Igor Podlubny. Hardcover. Publisher: Academic Press; (October 1998) ISBN 0-12-558840-2

\item[[5]]
\label{Herrmann}
Fractional Calculus. An Introduction for Physicists, by Richard Herrmann. Hardcover. Publisher: World Scientific, Singapore; (February 2011) ISBN 978-981-4340-24-3

\item[[6]]
Transformation of Fractional Derivatives Under  Space Rotation. By Ehab Malkawi and Akram A. Rousan. Published in International Journal of Applied Mathematics, Volume 16 No. 2, 175-185, 2004.

\end{itemize}

\end{document}